\pdfoutput=1
\documentclass[a4paper,12pt,english]{article}

\usepackage[pdftex]{graphicx}
\usepackage{graphicx}
\usepackage{colordvi}
\usepackage{tikz}
\usepackage{fancybox}
\usepackage{cancel}
\usepackage{amsmath}
\usepackage{slashed}
\usepackage{amssymb}
\usepackage{caption}
\usepackage{appendix}
\usepackage{hyperref} 
\usepackage{multirow}
\hypersetup{colorlinks=true, linkcolor=black, urlcolor=blue}
\usepackage{float}
\usepackage{cite}
\usepackage{mathtools}
\usepackage{bbm}

\usepackage{wasysym}

\usepackage{color}
\definecolor{gray}{rgb}{0.5,0.5,0.5}

\newcommand\lsim{\mathrel{\rlap{\lower4pt\hbox{\hskip1pt$\sim$}}
    \raise1pt\hbox{$<$}}}
\newcommand\gsim{\mathrel{\rlap{\lower4pt\hbox{\hskip1pt$\sim$}}
    \raise1pt\hbox{$>$}}}

\newcommand{\beq}{\begin{equation}}
\newcommand{\eeq}{\end{equation}}
\newcommand{\bea}{\begin{eqnarray}}
\newcommand{\eea}{\end{eqnarray}}
\newcommand{\bem}{\begin{pmatrix}}
\newcommand{\eem}{\end{pmatrix}}
\newcommand{\noi}{\noindent}
\newcommand{\non}{\nonumber}

\newcommand{\bet}{\begin{itemize}}
\newcommand{\eet}{\end{itemize}}
\newcommand{\ben}{\begin{enumerate}}
\newcommand{\een}{\end{enumerate}}

\begin{document}

\numberwithin{equation}{section}

\begin{flushright}
\end{flushright}

\bigskip

\begin{center}

{\Large\bf  UV finite GUT with SUSY Breaking}

\vspace{1cm}

\centerline{Borut Bajc$^{a\ }$\footnote{\href{mailto:borut.bajc@ijs.si}{borut.bajc@ijs.si}},  Manuel Del Piano$^{b,c\ }$\footnote{\href{mailto:manuel.delpiano-ssm@unina.it}{manuel.delpiano-ssm@unina.it}}
 and Francesco Sannino$^{b,c,d\ }$\footnote{\href{mailto:sannino@qtc.sdu.dk}{sannino@qtc.sdu.dk}}}

\vspace{0.5cm}
\centerline{$^{a}$ {\it\small J.\ Stefan Institute, 1000 Ljubljana, Slovenia}}
\centerline{$^{b}$ {\it\small Scuola Superiore Meridionale, Largo S. Marcellino, 10, 80138 Napoli NA, Italy}}
\centerline{$^{c}$ {\it \small INFN sezione di Napoli and Dipartimento di Fisica, E. Pancini, Univ. di Napoli, Federico II.}}
\centerline{\it \small  Complesso Universitario di Monte S. Angelo Edificio 6, via Cintia, 80126 Napoli, Italy}
\centerline{$^{d}$ {\it\small Quantum Theory Center ($\hbar$QTC) at IMADA and Danish IAS, University of Southern Denmark,  Denmark}} 
\end{center}

\bigskip

\begin{abstract}
We provide an example of an ultraviolet finite supersymmetric grand unified theory of safe rather than free nature endowed with a supersymmetric dynamical breaking mechanism. Our results simultaneously enlarge the number of ultraviolet consistent supersymmetric grand unified theories while providing a relevant example of how to achieve a consistent ultraviolet safe extension of the Standard Model enjoying the benefits of grand unified theories.

\end{abstract}


\newpage 


\section{Introduction }
Grand unified theories (GUT)s~\cite{Georgi:1974sy,Pati:1974yy} constitute one of the main guiding principles to construct extensions of the Standard Model (SM) with supersymmetric versions  \cite{Dimopoulos:1981zb} being among the most celebrated examples because they more naturally fit into the unification paradigm. However, successful models of supersymmetric (SUSY) grand unification typically feature a large number of matter fields that modifies the UV character of the GUT with a loss of asymptotic freedom and the appearance of Landau poles below the Planck scale as reviewed in \cite{Bajc:2016efj}.  One can envision  two ways to amend this issue. The first is to hope that (super) gravity itself could eventually intervene and rescue the high energy behaviour of  the theory. However, in this case the loss of asymptotic freedom is so severe that the resulting Landau pole  occurs at energies that are orders of magnitudes lower than the scale where quantum gravity could be able to influence the theory.  A second way to address  the original problem is to search for UV finite GUTs where asymptotic freedom is replaced by  interacting UV fixed points \cite{Bajc:2016efj}.   Following S. Weinberg \cite{Weinberg} achieving UV finiteness via interacting fixed points can also be referred to as asymptotically safe (or more simply safe) theories\footnote{This means {\it saving } them from the occurrence of Landau poles.}. The existence of four-dimensional gauge theories featuring fermions and scalars was established in \cite{Litim:2014uca} along with the stability of their ground state \cite{Litim:2015iea}. We will concentrate here on the super safe GUT possibility and associated SUSY breaking in order to realize the Standard Model at low energies. Non-supersymmetric GUTs  with interacting UV fixed points were investigated in \cite{Molinaro:2018kjz,Fabbrichesi:2020svm} and for extra dimensions in \cite{Cacciapaglia:2023ghp}. 

Supersymmetric non-perturbative fixed points of safe nature were investigated in \cite{Martin:2000cr,Intriligator:2015xxa,Bajc:2016efj} via a number of mathematical tools ranging from the $a$-maximization~\cite{Intriligator:2003jj} to unitary constraints \cite{Mack:1975je,Grinstein:2008qk} and positivity bounds \cite{Anselmi:1997am,Anselmi:1997ys}. Using these methods we discovered in  \cite{Bajc:2016efj} one popular supersymmetric GUT that can develop safety in the UV. This  is the renormalizable  SUSY SO(10)~\cite{Clark:1982ai,Aulakh:1982sw,Aulakh:2003kg} theory, which we will use as starting point of our analysis.  The novel part of our work consists in making the model phenomenologically viable by adding a SUSY breaking mechanism compatible with the safe nature of the theory.  

The plan of the work is the following. In Section~\ref{summaryofthemodel} we summarize the results of~\cite{Bajc:2016efj} and present the safe SUSY GUT model we will be using for the rest of the work. We then move to the introduction of the SUSY breaking sector in Section~\ref{susybreaking} following the radiative paradigm~\cite{Witten:1981kv,Dimopoulos:1997ww}. We first show that it is first needed to break  SO(10) to SU(5) and then break supersymmetry radiatively following~\cite{Bajc:2008vk}. We further show in Section~\ref{BbS} that adding the SUSY breaking sector is compatible with the overall safety of the model.  We finally provide our conclusions in Section~\ref{conclusions}. 

\section{Safe SO(10) summary}
\label{summaryofthemodel}

The model we have in mind is renormalisable and is thus necessarily composed of large representations~\cite{Clark:1982ai,Aulakh:1982sw,Aulakh:2003kg}. 
It consists of three copies of matter $16$-dimensional spinorial representations, plus the Higgs sector 
made of the representations $10$, $126$, $\overline{126}$ and $210$. The renormalisable Higgs 
superpotential is

\beq
W_H=m_{10}\,10^2+m_{210}\,210^2+m_{126}\,126\,\overline{126}+
\lambda\,210^3 + 
\alpha\,210\,10\,126+\beta\,210\,10\,\overline{126} + \gamma\,210\,126\,\overline{126}
\eeq

\noi
while the Yukawa part of the superpotential is

\beq
\label{WYUK}
W_{Yukawa}=16_i\left(Y_{10}10+Y_{126}\overline{126}\right)_{ij}16_j
\eeq

\noi
where $i,j=1,\ldots3$ are generation indices.This model has been called minimal for some time, 
until it was shown not in accord with experiment due to too small neutrino masses \cite{Aulakh:2005bd,Bajc:2005qe,Aulakh:2005mw,Bertolini:2006pe}. We will 
not be interested in this detail, being this model for us just a toy example of a UV safe theory. 
That this theory has a consistent candidate for a non-trivial UV fixed point has been shown in \cite{Bajc:2016efj}. 

To summarise this result, such a fixed point has the $R$-charges of all the fields in the UV fixed point 
equal to $2/3$ except for $16_1$ with a huge $R=113/6$. Clearly this is possible only providing 
the first generation of matter fields to be decoupled from the rest, i.e. for a superpotential (\ref{WYUK}) 
with $i,j$ running only from $2$ to $3$. This is obviously unrealistic, and we will assume that this can be 
somehow taken care of.  In any case we will not dwell into this problem further here and postpone its 
solution to a potential future publication. 

What we are interested here is the supersymmetry breaking 
part which was missed in \cite{Bajc:2016efj}. In fact we need to break supersymmetry sooner or later, and the 
way we do it could influence and even spoil the existence of the UV fixed point found in \cite{Bajc:2016efj}. 
This is studied in the next section: we will first show that the SO(10) model cannot break supersymmetry, 
find out that the SU(5) subgroup of the SO(10) model can do it, and finally show that the 
supersymmetry breaking sector only slightly changes but not destroys the UV fixed point.

\section{Breaking supersymmetry safely } 
\label{susybreaking}
To construct a realistic model we must break supersymmetry while preserving the safe nature of the original GUT theory. Because we are insisting on a self-consistent field theoretical approach we will concentrate on the class of supersymmetry breaking known as gauge mediation~\cite{Giudice:1998bp} following closely reference~\cite{Bajc:2008vk}. According to this paradigm supersymmetry breaking occurs radiatively in a distinct sector of the theory which develops a nonsupersymmetric metastable vacuum. Whether this metastable local minimum of the full potential exists depends on the details of the model. We have  the further restriction that the model must still be overall safe in the UV. 
 
 We will therefore first investigate whether SO(10) dynamics can lead to radiative supersymmetry breaking and show that the large number of matter fields hampers the result. Hence, we are led to study the case in which SO(10) spontaneously breaks first to SU(5) and then the latter radiatively breaks supersymmetry and show that the model is viable. 
 
 Before diving into the details of the specific models relevant for this work we sketch the radiative susy breaking scenario. In its minimal version the mechanism employs two non-singlet superfields $\Phi_1$ and $\Phi_2$ with the following gauge-invariant superpotential. 
 \begin{equation}
 W_{SB} = \mu \, \Phi_1  \Phi_2 +  \lambda \, \Phi_1^2  \Phi_2 \ .
 \end{equation}
The susy preserving global minimum occurs for $\Phi_2 = 0$ and $\Phi_1$ as solution of $\partial W_{SB}/ \partial \Phi_2 = 0$. Another possible local minimum can appear, depending on the details of the theory, for $\Phi_1$ being a solution of $\partial W_{SB}/ \partial \Phi_1 = 0$ and $\Phi_2$ a classical flat direction. The latter is the metastable vacuum we are interested in, provided it is not a maximum in the $\Phi_2$ direction at the quantum level. Specifically, following reference \cite{Witten:1981kv,Dimopoulos:1997ww,Bajc:2008vk}, the potential evaluated at the susy breaking extremum can be approximated to be: 
\beq
V\approx\frac{|F_2|^2}{Z_2} \ ,
\label{VSB}
\eeq
with $F_2$ the auxiliary component of the $\Phi_2$ superfield evaluated at the extremum and $Z_2$  the square of its wave function renormalization. Clearly,  different models predict distinct quantum corrections which we shall compute in the examples below.

\subsection{SO(10) with two 54s}
 To apply the susy breaking mechanism we sketched above we introduce in SO(10) two non-gauge singlet fields that we adopt to construct the needed quadratic and cubic gauge invariant terms. The minimal dimension of the fields are then 54 to  be able to construct the following superpotential
\beq
\label{SBold}
W_{SB} = \mu \,(54_1)_{AB}(54_2)_{BA}+ \lambda\,(54_1)_{AB}(54_1)_{BC}(54_2)_{CA} \ ,
\eeq
where the indices $A,B$ and $C$ run from one to ten\footnote{For the model to be phenomenologically viable quartic operators should be added as shown in \cite{Bajc:2008vk}, which however don't affect what we are going to discuss below.  We will come back to this point in the next sections.}. 
To explicitly test the viability of this construction we need to compute the running of $Z_2$ which depend on the beta function for $\lambda$ and the SO(10) gauge coupling $g_{10}$ derived in the appendix. The relevant system of beta functions is: 
\bea
\frac{dg_{10}^2}{d\tau}&=&133g_{10}^4\\
\frac{d(\log{\lambda^2})}{d\tau}&=&-60g_{10}^2+28\lambda^2\\
\frac{d(\log{Z_2})}{d\tau}&=&20g_{10}^2-\frac{28}{5}\lambda^2
\eea
with  $\tau=\log{\mu}/(8\pi^2)$. To determine whether the potential at the non-supersymmetric extremum  in \eqref{VSB} is a minimum we need the first derivative of $Z_2$ to vanish and the second to be negative.  The first condition relates $\lambda$ and $g_{10}$ as follows: 
\beq
\lambda^2=\frac{20}{\frac{28}{5}}g_{10}^2 \ ,
\eeq
while the second condition 
\noi
is not satisfied since 
\beq
\frac{1}{Z_2}\frac{d^2Z_2}{d\tau^2}=20\left(133+60-\frac{28\times20}{\frac{28}{5}}\right)g_{10}^4
=1860\,g_{10}^4>0 \ .
\eeq
This shows that the nonsupersymmetric extremum is, in this case, a maximum. This occurs because the one-loop coefficient for $g_{10}$ has large contributions coming from the matter field content of the theory while the negative contribution stemming from the $\lambda$ coupling is insufficient to offset the gauge-coupling contribution. One could imagine to consider different representations for the susy breaking sector fields  $\Phi_{1,2}$ however, for SO(10), it would require consider even higher gauge-group representations. The latter would typically  increase the contribution to the coefficient of the gauge beta function in such a way that the maximum cannot be turned into a minimum.  {An explicit computation along the lines shown above confirms this expectation}. 

We know, however, that it is possible to break supersymmetry with the mechanism discussed above in SU(5) models of grand unification \cite{Bajc:2008vk}.  All we need to do is to allow for spontaneous breaking of SO(10) to SU(5) and enact susy breaking at this latter stage. Fortunately, such a breaking is allowed and has already been investigated in the literature \cite{Bajc:2004xe,Fukuyama:2004ps}.  To further reduce the number of degrees of freedom for the target SU(5) we add the following operator 

\beq
\label{45}
 W_{\eta}=\eta\, Tr\left(54_154_245\right) \ ,
\eeq
featuring a new 45 superfield that has the task, after spontaneous symmetry breaking, to give mass to the two 
$(15-\overline{15})$s inside the 54s leaving behind two light 24s of SU(5) needed to break supersymmetry\footnote{We stress that the presence of this term, being linear in the 45, does not affect the vacuum structure of the SO(10) model.}.   The original superpotential features now also terms involving this new 45, but it can be shown \cite{Fukuyama:2004ps}, that the vacuum still aligns in the SU(5) intact direction.

\noi 
 
 
 \subsection{Intermediate SU(5) with two 24} 
 We  are now left, at an intermediate energy, with an SU(5) model featuring two extra light 24 superfields that, following reference  \cite{Bajc:2008vk}, can be used to break supersymmetry dynamically with the  part of the superpotential relevant for computing the contributions to the relevant beta functions that reads: 
\beq
W_{SB}=\lambda Tr\left(24_224_1^2\right) \ ,
\eeq

\noi
yielding\footnote{The difference in the one-loop coefficients with respect to reference \cite{Bajc:2008vk}  derives from the fact that we have extra 24s in the SU(5) spectrum. }. 
\bea
\frac{dg_5^2}{d\tau}&=&7g_5^4\\
\frac{d(\log{\lambda^2})}{d\tau}&=&-30g_5^2+21\lambda^2\\
\frac{d(\log{Z_2})}{d\tau}&=&10g_5^2-\frac{21}{5}\lambda^2 \ .
\eea

\noi
Vanishing of the first derivative of $Z_2$ requires: 
\beq
\lambda^2=\frac{10}{\frac{21}{5}}g_5^2 \ ,
\eeq
while the value of the second  derivative is now: 
\beq
\label{d2Z2SU5}
\frac{1}{Z_2}\frac{d^2Z_2}{d\tau^2}=10\left(7+30-\frac{21\times10}{\frac{21}{5}}\right)g_5^4=-130g_5^4<0 \ .
\eeq

\noi
The result guarantees that we have now a local minimum of the effective potential and that therefore we can successfully beak supersymmetry. 
 
\begin{figure}[htb]
\begin{center}
\includegraphics[width=0.95\linewidth]{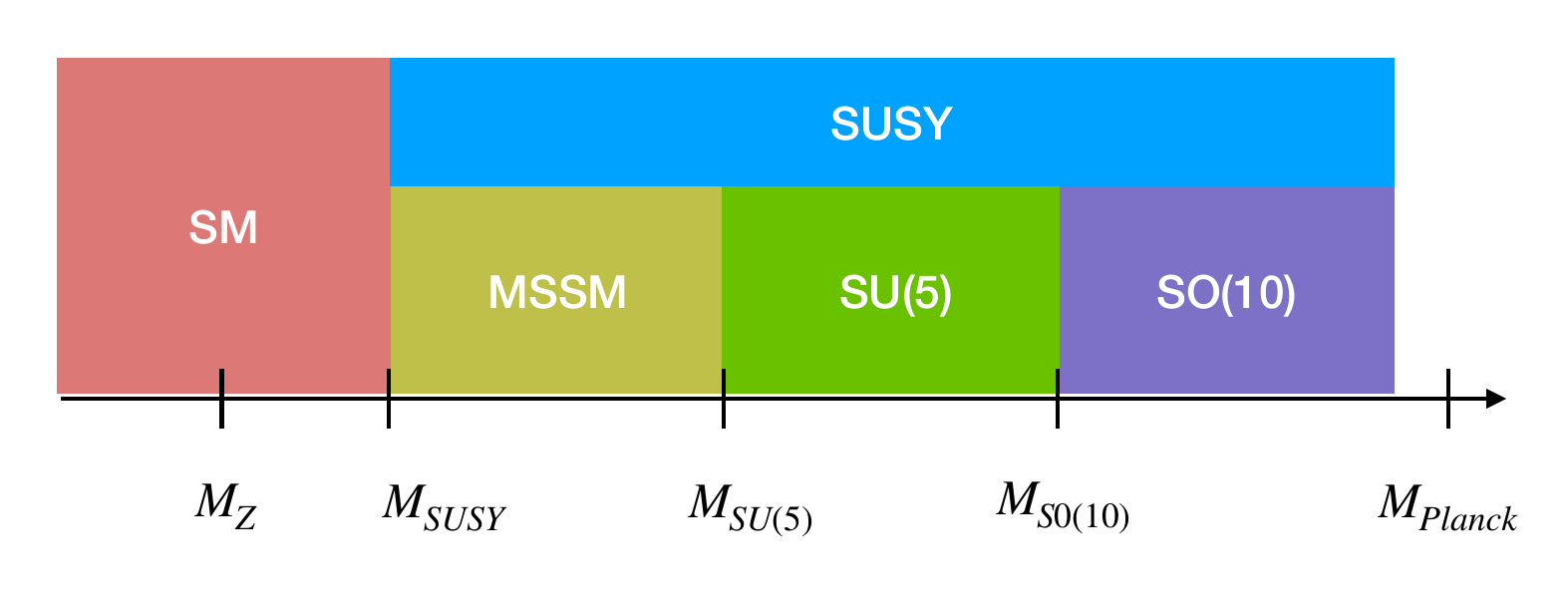}
\caption{Visualization of the relevant energy scales  and their ordering: The onset of the SO(10) UV finite theory occurs at $M_{SO(10)}$ and below this scale the theory  breaks to  SUSY SU(5). MSSM occurs below $M_{SU(5)}$  with soft masses appearing at $M_{SUSY}$ below which one recovers the Standard Model particle states. The SUSY breaking mechanism operates at $M_{SU(5)}$ generating the soft scale $M_{SUSY}$.    }
\label{figura}
\end{center}
\end{figure}

\section{Broken but safe}
 \label{BbS}
Here we check that the whole model, including the susy breaking sector, is still ultraviolet finite. To do so we  first extend 
$W_{SB}$ \eqref{SBold} to the more complete phenomenologically viable version  \cite{Bajc:2008vk}: 

\beq
\label{WOmega}
W_{\rm SB-complete}=Tr\left(-M\Omega_1\Omega_2+\Omega_1\left(\mu_1 54_2+\lambda_1 54_1 54_2\right)
+\Omega_2\left(\mu_2 54_1+\lambda_2 54_1^2\right)\right) \ ,
\eeq
where $\Omega_{1,2}$ are new $54$ heavy dimensional fields yielding the old $W_{SB}$ of  \eqref{SBold} upon being integrated out. We note that, while the original susy breaking potential is sufficient to break supersymmetry it predicts light states upsetting the coupling unification.  This generalization above lifts the light states to the grand unified scale thereby preserving the coupling unification \cite{Bajc:2008vk}. 

The model that needs to be safe compared to the original one \cite{Bajc:2016efj} that motivated this work contains now four more $54$ superfields, one more $45$ and the addition of the two superpotentials $W_{\rm SB-complete}$ of \eqref{WOmega} and $W_{\eta}$ in \eqref{45}. Naturally, we have also to allow for the operators mixing the new $45$ with the original fields (210,126,$\overline{126}$ and 10). 
 
 Since mass operators are irrelevant in the UV it is sufficient to consider in the UV only the trilinear terms. Given the promiscuous nature of the $45$ superfield and the already established $R$ charge for the original Higgs superfields mentioned above, i.e. (210,126,$\overline{126}$ and 10) one can show that 
\beq
R(45)=R(\Omega_1)=\frac{2}{3} \ , 
\eeq
\noi
where the second identity comes from comparing \eqref{45} with \eqref{WOmega} yielding the following relations for the remaining superfields  
\beq
R(54_2)=\frac{4}{3}-R(54_1)\quad,\quad R(\Omega_2)=2-2R(54_1) \ .
\eeq

\noi
Employing  $a$-maximisation  for
\beq
a_1(R(54_1))+a_1(R(54_2))+a_1(R(\Omega_2))
\eeq

\noi
yields, in the UV, $R(54_2)=2/3$ 
Therefore the  overall modification to $\Delta a$  with respect to the theory without susy breaking of \cite{Bajc:2016efj} comes from the running of the couplings encoded in the NSVZ \cite{Novikov:1983uc} 
susy beta function and yields: 
\beq
R(16_1)=\frac{169}{6}\quad,\quad \Delta a=961950
\eeq
\noi
compared to  $R(16_1)=113/6$ and $\Delta a=2447486/9$ of the susy unbroken case in \cite{Bajc:2016efj}. Overall, our result shows that the susy breaking sector can naturally be included in a UV safe theory without profoundly upsetting its nature.

\section{\label{conclusions}Conclusions}
We  constructed  a working example of an ultraviolet super conformal SO(10)  grand unified theory that  dynamically breaks supersymmetry down to the Standard Model  \cite{Bajc:2016efj}.  We first observed that SO(10) cannot break supersymmetry dynamically because its quantum corrected potential does not feature a non-supersymmetric local minimum and then we showed that spontaneously breaking SO(10) to supersymmetric SU(5)  allow us to use the latter dynamics to break supersymmetry.  We summarise in figure~\ref{figura}  the relevant energy scales  and their ordering.

 We have therefore shown that one can now employ wider classes of phenomenologically relevant supersymmetric grand unified theories allowing for ultraviolet interacting non-perturbative fixed points dynamically breaking supersymmetry at intermediate energies with the Standard Model in the infrared.  In the future it would be interesting to investigate the E6 super grand unified theory and more generally the overall approach embedding into superstrings.

 \subsubsection*{Acknowledgments}
BB thanks the Quantum Theory Center at IMADA and the Danish Institute for Advanced Study of the University of Southern Denmark, for their hospitality during which part of this work has been performed. BB is supported by the Slovenian Research Agency under the research core funding No. P1-0035 and in part by the research grant J1-4389. The work of FS is partially supported by the Carlsberg Foundation, grant CF22-0922.


\end{document}